\documentclass[9pt,twocolumn,twoside]{optica_article}
\journal{opticajournal} 

\articletype{Research Article}

\usepackage[utf8]{inputenc}
\usepackage{textgreek}
\usepackage{siunitx}
\usepackage{multicol}

\newcommand{\dv}[2]{\frac{\mathrm{d}#1}{\mathrm{d}#2}}

\graphicspath{ {./images/} }

\begin{document}
\title{Compact, folded multi-pass cells for energy scaling of post-compression}

\author{Arthur Sch\"onberg,\authormark{1,*} Supriya Rajhans,\authormark{1,2} Esmerando Escoto,\authormark{1} Nikita Khodakovskiy,\authormark{1}  Victor Hariton,\authormark{1} Bonaventura Farace,\authormark{1} Kristjan Põder,\authormark{1} Ann-Kathrin Raab,\authormark{3} Saga Westerberg,\authormark{3} Mekan Merdanov,\authormark{3} Anne-Lise Viotti,\authormark{3} Cord L. Arnold,\authormark{3} Wim P. Leemans,\authormark{1} Ingmar Hartl,\authormark{1} and Christoph M. Heyl\authormark{1,4,5}}

\address{
\authormark{1}Deutsches Elektronen-Synchrotron DESY, Notkestraße 85, 22607 Hamburg, Germany\\
\authormark{2}Friedrich-Schiller-Universitat Jena, Max-Wien-Platz 1, 07743 Jena, Germany\\
\authormark{3}Department of Physics, Lund University, P.O. Box 118, SE-221 00 Lund, Sweden\\
\authormark{4}GSI Helmholtzzentrum f\"ur Schwerionenforschung GmbH,  Planckstraße 1, 64291 Darmstadt, Germany\\
\authormark{5}Helmholtz-Institute Jena, Fr\"obelstieg 3, 07743 Jena, Germany}

\email{\authormark{*}arthur.schoenberg@desy.de} 


\begin{abstract*} 
Combining high peak- and high average power has long been a key challenge of ultrafast laser technology, crucial for applications such as laser-plasma acceleration and strong-field physics. 
A promising solution lies in post-compressed ytterbium lasers, but scaling these to high pulse energies presents a major bottleneck.
Post-compression techniques, particularly Herriott-type multi-pass cells (MPCs), have enabled large peak power boosts at high average powers but their pulse energy acceptance reaches practical limits defined by setup size and coating damage threshold. 
In this work, we address this challenge and demonstrate a novel type of compact, energy-scalable MPC (CMPC). 
By employing a novel MPC configuration and folding the beam path, the CMPC introduces a new degree of freedom for downsizing the setup length, enabling compact setups even for large pulse energies. 
We experimentally and numerically verify the CMPC approach, demonstrating post-compression of \qty{8}{\milli\joule} pulses from \qty{1}{\pico\s} down to \qty{51}{\femto\s} in atmospheric air using a cell roughly \qty{45}{\cm} in length at low fluence values. 
Additionally, we discuss the potential for energy scaling up to \qty{200}{\milli\joule} with a setup size reaching \qty{2.5}{\m}.
Our work presents a new approach to high-energy post-compression, with up-scaling potential far beyond the demonstrated parameters. 
This opens new routes for achieving the high peak and average powers necessary for demanding applications of ultrafast lasers.
\end{abstract*}

\section{Introduction}
\label{Introduction}

Ultrafast laser technology has experienced immense progress within recent years. 
Ultrashort, high-peak power lasers are used in a vast range of applications, including attosecond science and high-harmonic generation \cite{Agostini2004,Corkum2007,Krausz2009,Heyl2012}, laser-plasma acceleration \cite{Esarey2009,Albert2021} or high-field science including laser-based nuclear fusion \cite{AbuShawareb2022}. However, developing a laser source which is simultaneously average and peak power scalable remains a major challenge.

The invention of mode-locked solid-state laser technology, in particular Titanium-doped sapphire (Ti:Sa) lasers in combination with chirped-pulse amplification (CPA) enabled ultrashort, few-cycle pulses with unprecedented pulse energy \cite{Spence1991,Backus1997,Fattahi2014}.
Nowadays, peak powers exceeding the Terawatt regime are routinely employed \cite{Backus1997}. 
While excelling in peak power performance, Ti:Sa amplifiers are commonly constrained in average power to a few tens of Watts, which can be attributed to their large quantum defect \cite{Wolter2017}. 
As an alternative to laser amplification in active gain media, optical parametric processes can be employed. 
In particular optical parametric chirped-pulse amplifiers (OPCPA) offer broad bandwidths supporting few-cycle pulses and simultaneously high average powers \cite{Fattahi2014,Witte2012}.
However, OPCPA systems suffer from low pump-to-signal efficiencies typically around 10-20\% for pulses in the range of 10s of femtoseconds (fs) \cite{Yin2016}. 
Ultrafast Ytterbium (Yb) -based laser architectures on the other hand provide excellent average power scalability exceeding \qty{10}{\kilo\watt} \cite{Mueller2020}, but pulse durations limited to 100s of femtoseconds up to about 1 picosecond (ps). Combining Yb lasers with efficient post-compression methods supporting large (>10) compression factors and high pulse energies can offer an excellent solution to the power scaling challenge. 
 
In recent years, a number of post-compression techniques have been developed, mostly relying on self-phase modulation (SPM) as the nonlinear process for spectral broadening \cite{Viotti2022}. 
In particular, gas-based technologies provide excellent tools for post-compression of high power lasers.
Example systems rely on gas-filled hollow-core fibers (HCF) \cite{Nagy2019,Travers2019,Travers2024}, cascaded focus and compression (CASCADE) \cite{Tsai2022}, white-light filaments \cite{Schulz2009}, as well as Herriott-type multi-pass cells (MPCs)\cite{Schulte2016,Weitenberg2017,Balla2020,Viotti2022}. 
In HCFs, post-compression of \qty{70}{\milli\joule} \qty{220}{\femto\second} pulses down to \qty{30}{\femto\second} has been demonstrated in a \qty{3}{\meter} long fiber \cite{Fan2021}. 
Post-compression of very high pulse energies in the multiple Joule range has been achieved via thin-film spectral broadening techniques. However, typical compression factors lie in the range of only 2-5 \cite{Mourou2014, Bleotu2022}.
Similar to HCFs, MPCs enable large compression factors reaching 10-20 or more while supporting a wide range of pulse energies. In addition, MPCs support high average powers \cite{Mueller2021} and outperform HCFs in system footprint especially for large compression factors  \cite{Viotti2022}. 
The maximum attainable energy acceptance in a standard, two-mirror MPC is directly proportional to its size \cite{Heyl2016,Viotti2022}. A record of \qty{200}{\milli\joule} has been achieved in a \qty{10}{\meter} long MPC setup \cite{Pfaff2023}.
Further energy up-scaling leads to MPC sizes that are impractical for standard laboratory settings. 
The development of a highly efficient post-compression method supporting large compression factors and high pulse energies thus remains a key challenge. 

We here introduce a new MPC type, the compact MPC (CMPC) which possesses a weakly focused fundamental mode as well as a linear beam pattern on the focusing mirrors. 
This geometry allows us to fold the beams inside the MPC using two additional planar mirrors, thus introducing a new energy scaling parameter, the folding ratio $\Gamma$. 
The CMPC in principle allows for an arbitrary amount of folding and thus, very compact setup sizes while sharing key properties of standard MPCs such as high average power support, excellent beam quality and efficiency. 
We experimentally demonstrate spectral broadening of \qty{1030}{\nano\meter}, \qty{8}{\milli\joule}, \qty{1}{\pico\second} pulses in a CMPC in atmospheric air using a setup with an effective length of around \qty{45}{\cm}. 
We keep the maximum mirror fluence at a moderate level of around \qty{170}{\milli\joule\per\centi\meter^2} and demonstrate compressibility of \qty{1}{\pico\s} input pulses down to \qty{51}{\femto\second} with an MPC throughput reaching 89\% while maintaining excellent spatio-temporal pulse characteristics.

\section{Concept}
\label{sec:concept}
\begin{figure}[t]
    \includegraphics[width=\textwidth]{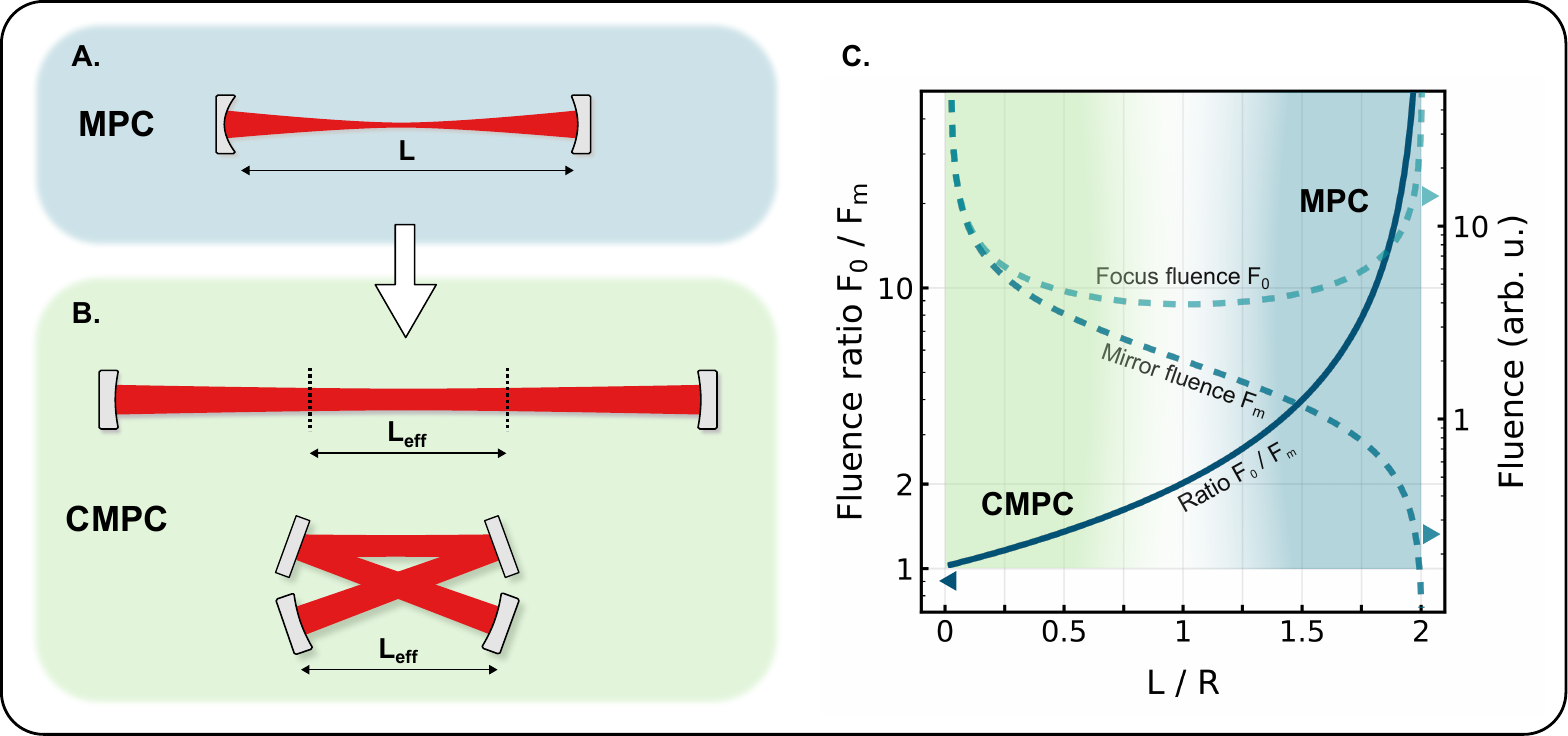}
    \centering
    \caption{MPC and CMPC schematic and configuration regime. (A) Standard MPC mode. (B) CMPC mode without and with folding. In the unfolded geometry the CMPC is typically longer than a comparable MPC, reducing to much shorter length $L_{\mathrm{eff}}$ after folding. (C) Typical configurations regimes and corresponding fluence characteristics of MPC and CMPC within the stability range.}
    \label{fig:MPC_CMPC_simple}
\end{figure}
Most MPCs demonstrated for post-compression to date rely on two identical concave mirrors, resembling the most basic optical cavity arrangement. However, more complex designs employing a convex and a concave mirror \cite{Omar2023,Hariton2023} or even multiple additional mirrors can provide advantageous mode-forming capabilities. 
MPCs with more than two mirrors have been proposed in previous works focusing on energy-scaling of MPCs \cite{Heyl2022,Hariton2023}. 
The concept of the CMPC is based on a weakly focused beam and folding of the beam path via multiple reflections on two additional, planar mirrors in each pass through the cell, as shown in Fig. \ref{fig:MPC_CMPC_simple}. 
This provides an additional tuning parameter, namely the folding ratio $\Gamma$, which reduces the length of the CMPC by $L_{\mathrm{eff}} \approx L/\Gamma$. 
Figures \ref{fig:MPC_CMPC_simple}(A) and \ref{fig:MPC_CMPC_simple}(B) depict the principle of beam folding and the effective size reduction of the cell together with the configuration regimes for standard MPCs and the CPMC.

\begin{figure}[t]
    \includegraphics[width=\textwidth]{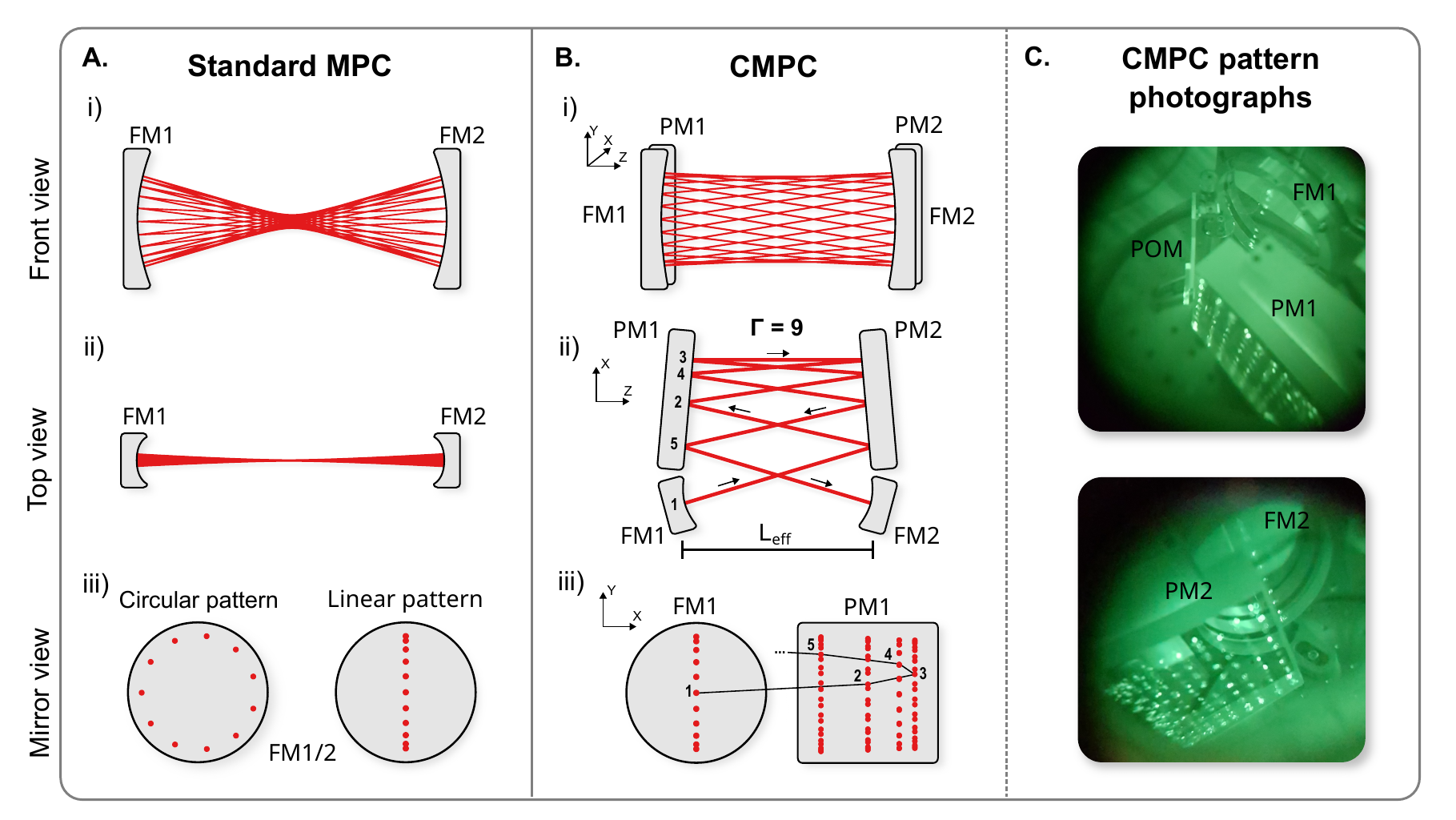}
    \centering
    \caption{Schematics of a standard MPC and CMPC. (A) Standard MPC geometry and beam pattern considering a linear beam pattern alignment (FM1/2: focusing mirrors). View of the beam path from the front (i) and from the top (ii). (iii) Resulting patterns for circular/linear alignment on FM1/2. (B) CMPC geometry and beam patterns. (i) Front view with additional planar mirrors (PM1/2) behind FM1/2. (ii) Top view onto the CMPC with beam folding ratio $\Gamma=9$. The numbers (1-5) are indicating the order of reflections of the first pass. (iii) Beam patterns on the mirrors FM1/2 and PM1/2. The black line follows the beam reflections 1-5 on the mirrors. (C) Photographs of experimental beam patterns with $N=11$, $k=3$ and $\Gamma=25$. POM: pick-off mirror.}
    \label{fig:MPC_CMPC}
    \end{figure}

In general, the geometry of a symmetric Herriott-type MPC - including a CMPC - is fully determined by three of the following four parameters: radius-of-curvature (ROC) of the mirrors $R$, the propagation length between the two focusing mirrors $L$, the number of round-trips $N$, and the configuration parameter $k$. These parameters are related by \cite{Viotti2022}:
\begin{equation}
    \frac{L}{R} = 1 - \cos{\left(\frac{\pi k}{N}\right)}\ .
    \label{eq:RLkN_formula}
\end{equation}
Typically, $k$ is chosen to be integer and relatively prime to the number of round-trips $N$ in order to ensure the re-entrant condition of the MPC. 
Moreover, the parameter $k$ determines the angular advance of the beam on the MPC mirror, which is equivalent to the Gouy phase per pass $\phi_{\text{G}}^{(1)} = \pi k/N$ \cite{Viotti2022}. 
The amount of Gouy phase per pass determines the caustic of the beam in the MPC. For a standard MPC, $\phi_{\text{G}}^{(1)}$ approaches $\pi$, which leads to a large mode-size $w_{\mathrm{m}}$ on the mirrors and a small waist $w_0$ in the focus. 
In the CMPC, the beam propagates within the Rayleigh-range, where the accumulated Gouy phase per pass is $\phi_{\text{G}}^{(1)}<\pi/2$.
This is the case when $L/R<1$, where the beam radius at the mirror $w_{\mathrm{m}}$ is comparable to the beam radius at focus $w_0$. 
The peak fluence of the beam follows a similar behavior, 
with $F_0/F_{\mathrm{m}} < 2$ for $L/R<1$ [Fig. \ref{fig:MPC_CMPC_simple}(C)].
This weak focusing geometry with small fluence variations enables the placement of additional optical components within the beam path of the CMPC and thus folding of the beam. In addition, a weakly focused mode eliminates pulse energy limitations arising due to ionization in standard gas-filled MPCs. 
Figure \ref{fig:MPC_CMPC} illustrates the CMPC scheme. In Fig. \ref{fig:MPC_CMPC}(A) the setup schematic of a standard MPC with linear pattern alignment is shown. 
Here, the strongly focused beams propagate directly between the two focusing mirrors FM1 and FM2, without any optical components in-between. 
In the CMPC [Fig. \ref{fig:MPC_CMPC}(B)], two additional planar mirrors PM1 and PM2 are placed behind FM1 and FM2. As seen in Fig. \ref{fig:MPC_CMPC}(B)(ii), the beam propagates from FM1 to PM2 and PM1, where it is folded multiple times before reaching FM2. 
Here, the depicted folding ratio is $\Gamma=9$. 
The total path length between FM1 and FM2 is the length determined by Eq. (\ref{eq:RLkN_formula}). 
However, the effective length is now reduced by $L_{\mathrm{eff}} \approx L/\Gamma$, which under the correct choice of $R$ and $\Gamma$ can become significantly shorter than a standard MPC with similar pulse energy acceptance.

The pulse energy acceptance of a gas-filled MPC can be estimated considering the mirror fluence and the focus intensity. 
The beam waist radius in the focus and on the mirrors in an MPC can be calculated using the equations $w_0^2 = R\lambda/2\pi\,\sin{\left(\pi /kN\right)}$ and  $w_m^2 = R\lambda/\pi\,\tan{\left(\pi k/2N\right)}$ respectively, where $\lambda$ is the wavelength of the laser \cite{Heyl2022}. The parameters $L/R$ can be directly mapped to $k/N$ according to Eq. (\ref{eq:RLkN_formula}). 
Assuming a laser with Gaussian pulses and pulse energy $E$, the maximum peak fluence in the focus $F_0 = 2E/\pi w_0^2 $, as well as the peak fluence on the mirror $F_{\mathrm{m}} = 2E/\pi w_m^2$ can be calculated \cite{Heyl2022}.
For a standard MPC consisting of two identical focusing mirrors, the fluence on these mirrors should not exceed the threshold fluence $F_{\text{th}}$, yielding a  maximum pulse energy
\begin{align}
    E_{\mathrm{MPC}} \leq \frac{R \lambda F_{\text{th}} }{2}\,\tan{\left(\pi k/2N\right)} \, .
    \label{eq:EmaxMPC}
\end{align}
For simplicity, the second energy limitation which can arise at high focus intensity due to ionization is omitted in this discussion as it has little or no relevance for the CMPC. 
In a CMPC, where beam folding is achieved by additional mirrors [Fig. \ref{fig:MPC_CMPC}] the beam can be reflected at almost any point between the focusing mirrors including the focus. Therefore, the condition for avoiding damage needs to be $F_0 \leq F_{\text{th}}$ and thus
\begin{align}
    E_{\text{CMPC}} \leq \frac{R \lambda F_{\text{th}} }{4}\,\sin{\left(\pi k/N\right)}\ .
    \label{eq:EmaxCMPC}
\end{align}
In addition, replacing $R$ with $L$ using Eq. (\ref{eq:RLkN_formula}) and writing $L$ in terms of $L_{\mathrm{eff}} \approx L/\Gamma$, we obtain:
\begin{align}
    E_{\text{CMPC}} \leq \,\frac{\lambda \, F_{\text{th}} \, \Gamma \, L_{\text{eff}}}{4 \, \tan{(\pi k/2N)}} \, \ .
    \label{eq:EmaxCMPC2}
\end{align}
\begin{figure}[htb]
    \includegraphics[width=0.55\textwidth]{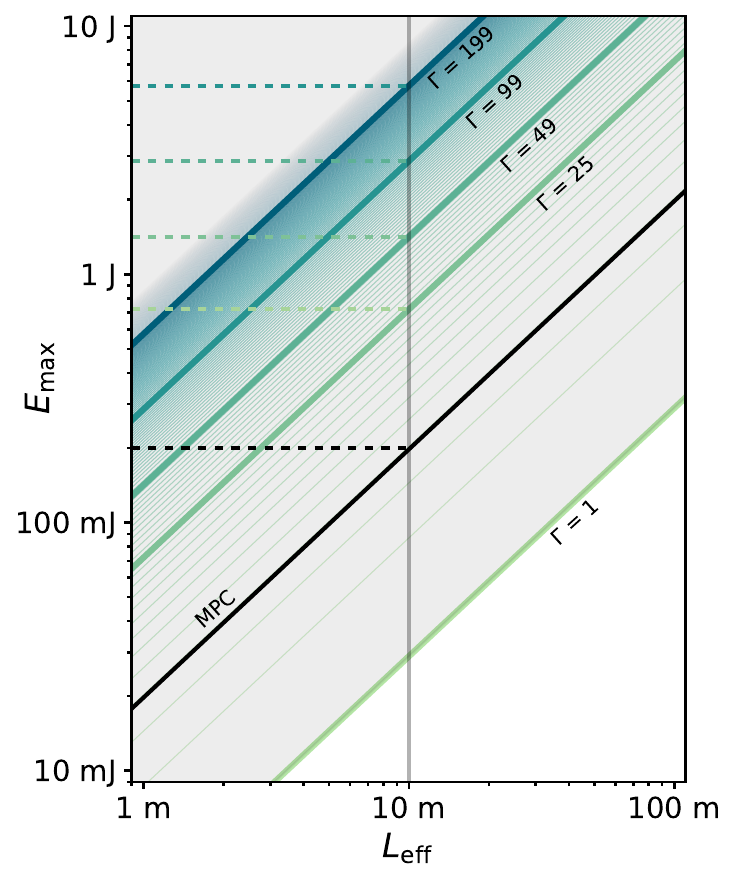}
    \centering
    \caption{Energy scaling characteristics: Calculated pulse energy acceptance $E_{\mathrm{max}}$ as a function of effective length $L_{\mathrm{eff}}$. The solid black line indicates a standard MPC as reported in reference \cite{Pfaff2023} [Eq. (\ref{eq:EmaxMPC})]. Each colored solid line represents a CMPC with a folding ratio $\Gamma$ [Eq. (\ref{eq:EmaxCMPC2})]. The gray line indicates the point of effective length of \qty{10}{\m}. The dashed lines show the corresponding pulse energies for selected values of $\Gamma$. We set the threshold fluence to $F_{\mathrm{th}}=\qty{0.5}{\joule\per\cm^2}$.}
    \label{fig:FE_kapL}
\end{figure}
Equations (\ref{eq:EmaxMPC})-(\ref{eq:EmaxCMPC2}) illustrate the energy scaling possibilities of MPCs. For standard MPCs [Eq. (\ref{eq:EmaxMPC})], the energy scaling options are restricted to maximizing the ratio $k/N \rightarrow 1$, increasing the ROC of the mirrors $R$ (which increases $L$ proportionally at constant $k/N$) or increasing the threshold fluence $F_{\text{th}}$.
In the case of CMPCs, energy scaling can be achieved differently. Since $k/N$ is generally smaller, the energy acceptance for the same set of parameters $R$ and $F_{\mathrm{th}}$ is typically reduced according to Eq. (\ref{eq:EmaxCMPC}). 
Thus, the first step of scaling is achieved by matching $R$ in Eq. (\ref{eq:EmaxCMPC}) such that $E_{\text{CMPC}} = E_{\text{MPC}}$.
This typically increases $L$, i.e. $L_{\text{CMPC}} > L_{\text{MPC}}$.
The second step is to fold the beam by the factor $\Gamma$ and decrease the length of the system by exploiting $L_{\text{eff}} = L/\Gamma$.

To illustrate the discussion with an example, we use a typical set of parameters for a standard MPC with $R=\qty{1}{\m}$ and a $\lambda = \qty{1030}{\nano\m}$, $\qty{1}{\pico\s}$ level laser.
The threshold fluence for quarter-wave stack high-reflectance dielectric coatings can e.g. be set to $F_{\mathrm{th}} = \qty{0.2}{\joule\per\cm^2}$, leaving headroom to damage. 
With $N=15$ round-trips and $k=14$, we arrive at $E_{\text{MPC}}\leq \qty{9.8}{\milli\joule}$ and a length of close to $\qty{2}{\m}$.
To match the pulse energy acceptance of the CMPC to the MPC, we first set $k=4$, increase the ROC to $R=\qty{25}{\m}$ and the length to $L=\qty{8.27}{\m}$. 
Now folding the beam $\Gamma=25$ times, we arrive at $E_{\text{CMPC}} \leq \qty{9.54}{\milli\joule}$ with an effective length of $L_{\text{eff}} = \qty{33}{\cm}$. 
A second example is illustrated in Fig. \ref{fig:FE_kapL}. 
Here we refer to the works reported by Pfaff et al., where \qty{200}{\milli\joule} pulses have been compressed in a \qty{10}{\m} long MPC, with a fluence on the mirrors of approximately \qty{0.5}{\joule\per\cm^2} \cite{Pfaff2023}. 
Figure \ref{fig:FE_kapL} shows that the energy acceptance increases with larger folding ratio and effective system length.
At 10 m length, a folding ratio of $\Gamma=25$ supports an energy of $\qty{700}{\milli\joule}$ and at $\Gamma=49$, the accepted energy exceeds 1 Joule. 
Conversely, for a pulse energy of \qty{200}{\milli\joule}, the CMPC size can be scaled down to about \qty{3}{\m} or \qty{1.5}{\m} considering $\Gamma=25$ and $\Gamma=49$, respectively.
As visible in Figure \ref{fig:FE_kapL}, for $\Gamma=1$, the fully "unfolded" CMPC is generally longer than the standard MPC considering an identical threshold fluence for both cases. This is due to the fact that the CMPC operates in the $L/R<1$ regime while the MPC operates at $L/R \approx 2$.

\section{Experiment}

\begin{figure}[t]
    \includegraphics[width=0.8\textwidth]{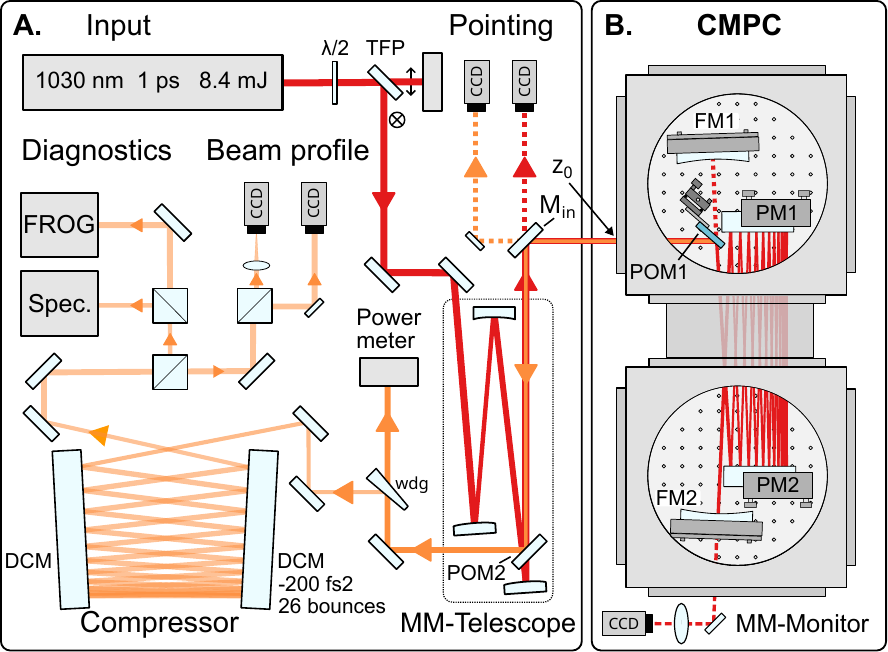}
    \centering
    \caption{Experimental setup. The CMPC input beam is indicated in red, the output beam in orange. FM: focusing mirrors. PM: planar mirrors. POM: pick-off-mirrors. $\mathrm{M}_{\mathrm{in}}$: input/output mirror. MM: mode-matching. DCM: double-chirped mirrors. CCD: cameras. $z_0$ reference point for stabilization measurements. TFP: thin-film polarizer. wdg: wedge. (A) Beam input, telescope and diagnostics used in the experiment. (B) CMPC setup with vacuum chamber.}
    \label{fig:setup}
\end{figure}
Our experimental setup displayed in Fig. \ref{fig:setup} uses a commercial innoslab Yb laser system delivering \qty{1030}{\nano\meter} \qty{1}{\pico\second} pulses at a repetition rate of \qty{1}{\kilo\hertz} with a pulse energy of > \qty{8}{\milli\joule}. 
After suitable mode-matching, the beam is sent onto the pick-off mirror POM1 and into the CMPC [Fig. \ref{fig:setup}(B)]. 
The CMPC consists of two concave 4 inch mirrors with $R=\qty{25}{\m}$ and two planar folding mirrors with size $\qty{10}{\cm}\times\qty{10}{\cm}$. 
The configuration is set to $N=11$ round-trips and $k=3$, which corresponds, according to Eq. (\ref{eq:RLkN_formula}) to an unfolded MPC length of $L=\qty{8.63}{\m}$. 
Following in-coupling at the pick-off mirror POM1, the beam is sent onto the planar folding mirror PM2 and is subsequently reflected by PM1, with both mirrors being aligned in a V-shaped configuration and separated by about $\qty{34}{\cm}$.
The beam continues its path as described in Fig. \ref{fig:MPC_CMPC}, forming the typical CMPC pattern. 
A folding ratio of $\Gamma=25$ results in an effective CMPC length of $L_{\mathrm{eff}}=\qty{34.5}{\cm}$. 
However, due to spatial constraints, the two focusing mirrors FM1 and FM2 are placed a few centimeters behind PM1 and PM2, making the total CMPC length slightly larger, resulting in around \qty{45}{\cm} [see Fig. \ref{fig:setup}(B)].
After propagating $N=11$ round-trips (22 passes), the beam returns on PM1 and is coupled out with a different angle in the vertical direction. 
The total number of mirror reflections amounts to $2N\Gamma = 550$ and the total propagation length is roughly \qty{190}{\m}. 
We use ambient air at atmospheric pressure (\qty{1006}{\milli\bar}) as the nonlinear medium for spectral broadening. 
Nevertheless, a chamber or housing is necessary to avoid beam fluctuations due to turbulences in air.
After out-coupling, the spectrally broadened beam is picked-off with another pick-off mirror POM2 [Fig. \ref{fig:setup}(A)]. 
Here, we measure an output power of around \qty{7.1}{\milli\watt} and thus a total transmission of roughly 89\%.
This corresponds to a reflectance of the mirrors of at least 99.98\% per reflection, disregarding clipping losses.
Subsequently, a wedge is used for beam sampling, reflecting around 8\%, corresponding to roughly \qty{570}{\milli\watt}. 
The transmission of the wedge is sent onto a power meter.
The wedge-reflected pulse is compressed using a chirped-mirror compressor with 26 reflections corresponding to about \qty{-5200}{\femto\s^2} of compensated second order dispersion.
\begin{figure}[t]
    \centering
    \includegraphics[width=0.55\textwidth]{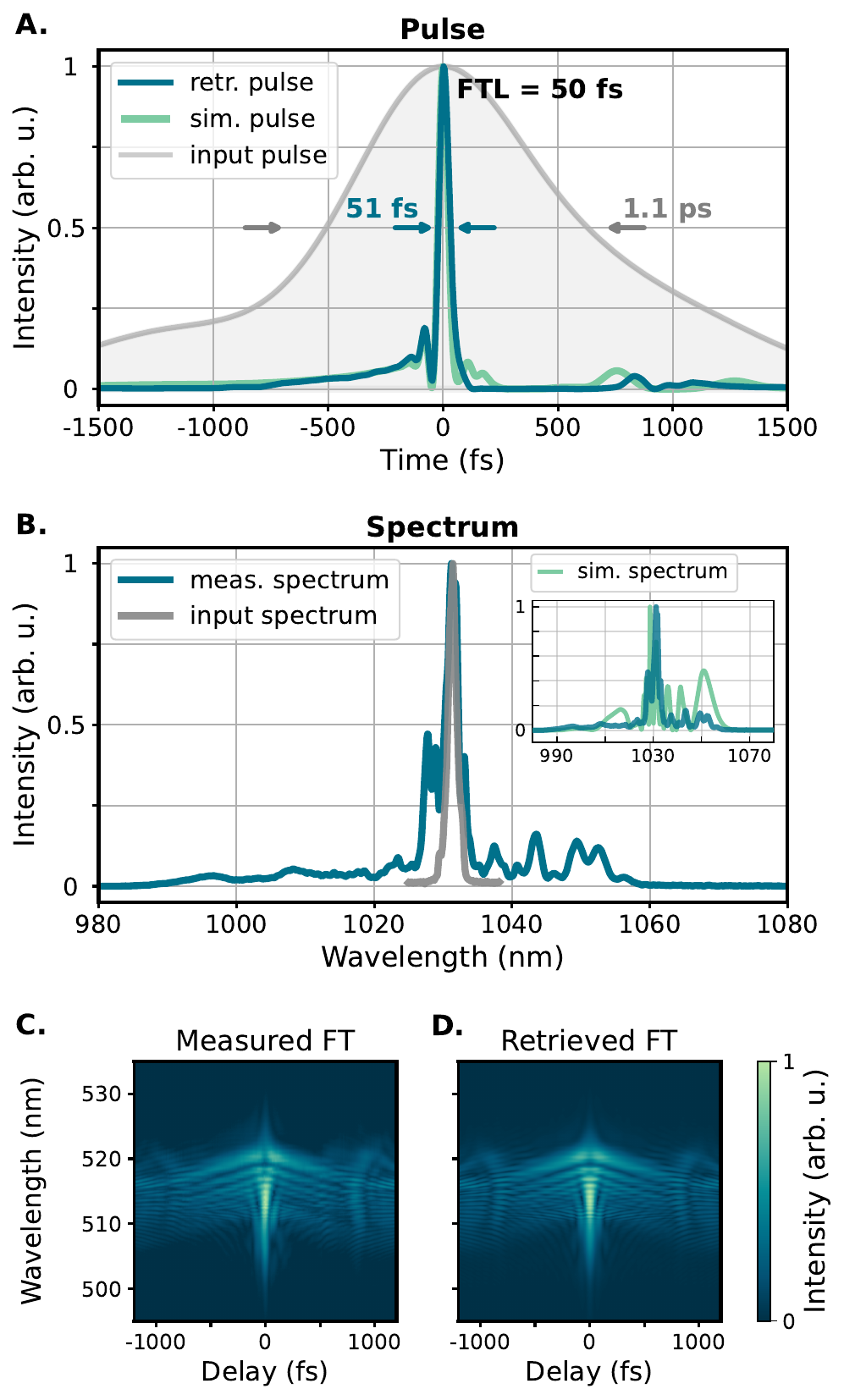}
    \caption{Measured and simulated pulse characteristics after CMPC spectral broadening and subsequent compression. (A) Reconstructed input and output temporal pulse. (B) Measured input spectrum and broadened spectrum. The inset shows the measured broadened spectrum along with simulated spectrum. (C) Measured FROG trace (FT). (D) Retrieved FROG trace (FROG error = 0.73\%). }
    \label{fig:pulse_spectrum_results}
\end{figure}
We analyze the compressed pulses using frequency-resolved optical gating (FROG) and measure the spectrum. 
In addition, the beam quality parameter $\mathrm{M}^2$ is analyzed using the same wedge reflection. 
We furthermore record input and output near-field pointing using the input/output mirror ($\mathrm{M}_{\mathrm{in}}$) transmission of both input and output beams. 
For direct comparison of input and output, both cameras are placed in the exact same distance $d=\qty{42}{\cm}$ from a common reference point $z_0$, located at the vacuum chamber input window.
\begin{figure}[t]
    \centering
    \includegraphics[width=0.55\textwidth]{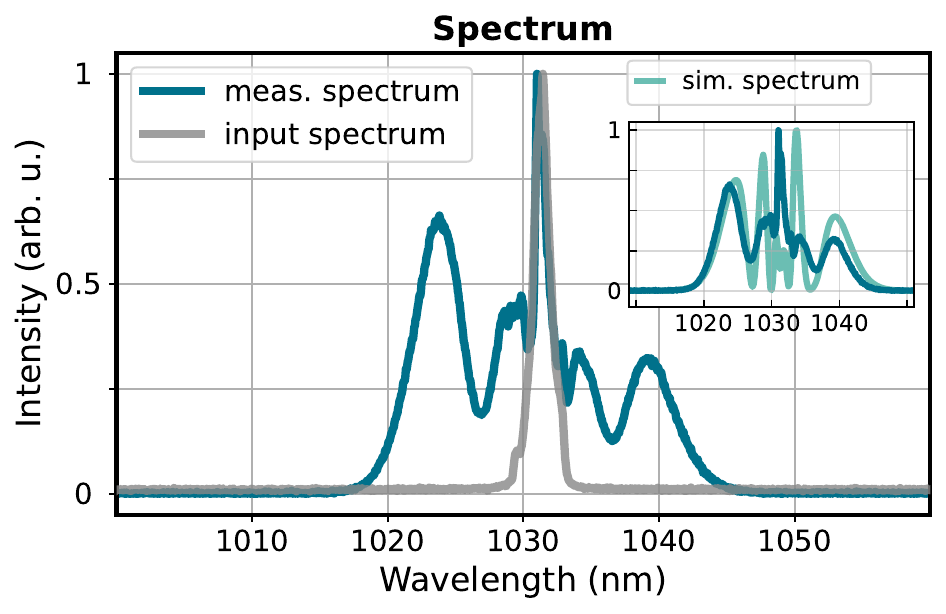}
    \caption{Spectral broadening of \qty{6.5}{\milli\joule} pulses in 1 bar of Krypton. CMPC input (gray line) and measured output (dark blue line) spectra are shown. The inset shows the simulated spectrum along with the measured output spectrum.}
    \label{fig:spectrum_results_Kr}
\end{figure}
The spectrum and FROG results are shown in Fig. \ref{fig:pulse_spectrum_results}. The input pulse has a pulse duration of about \qty{1.1}{\pico\s} with a Fourier-transform-limit (FTL) of about \qty{1}{\pico\s}. 
After spectral broadening in the CMPC, the output pulse FTL reaches \qty{50}{\femto\s}. Following compression, we reach \qty{51}{\femto\s} [Fig. \ref{fig:pulse_spectrum_results}(A)]. 
We simulate the spectral broadening process using the measured input pulse [Fig. \ref{fig:pulse_spectrum_results}(A)]. 
The simulations are conducted using our in-house developed (2+1)D radially symmetric simulation code based on Hankel-transforms in the spatial dimension, which is described in the supplemental document [Eqs. (\ref{eq:nl_eq})-(\ref{eq:Pnl2})].
The simulated output pulse agrees very well with the measured output pulse, exhibiting similar temporal characteristics after compression. In both measurement and simulation, we observe a temporal pedestal likely stemming from uncompressed spectral components in the longer wavelength range.
On the trailing edge, a post-pulse appears at around $800\,\mathrm{fs}$, which we attribute to a pulse-breakup caused by the delayed response of the Raman-Kerr contribution in air.
The measured and simulated spectra [Fig. \ref{fig:pulse_spectrum_results}(B)] also show similar characteristics as well as similar broadening, except for the strength of the side lobes compared to the background close to \qty{1030}{\nano\m}. 
\begin{figure}[h]
    \centering
    \includegraphics[width=0.6\textwidth]{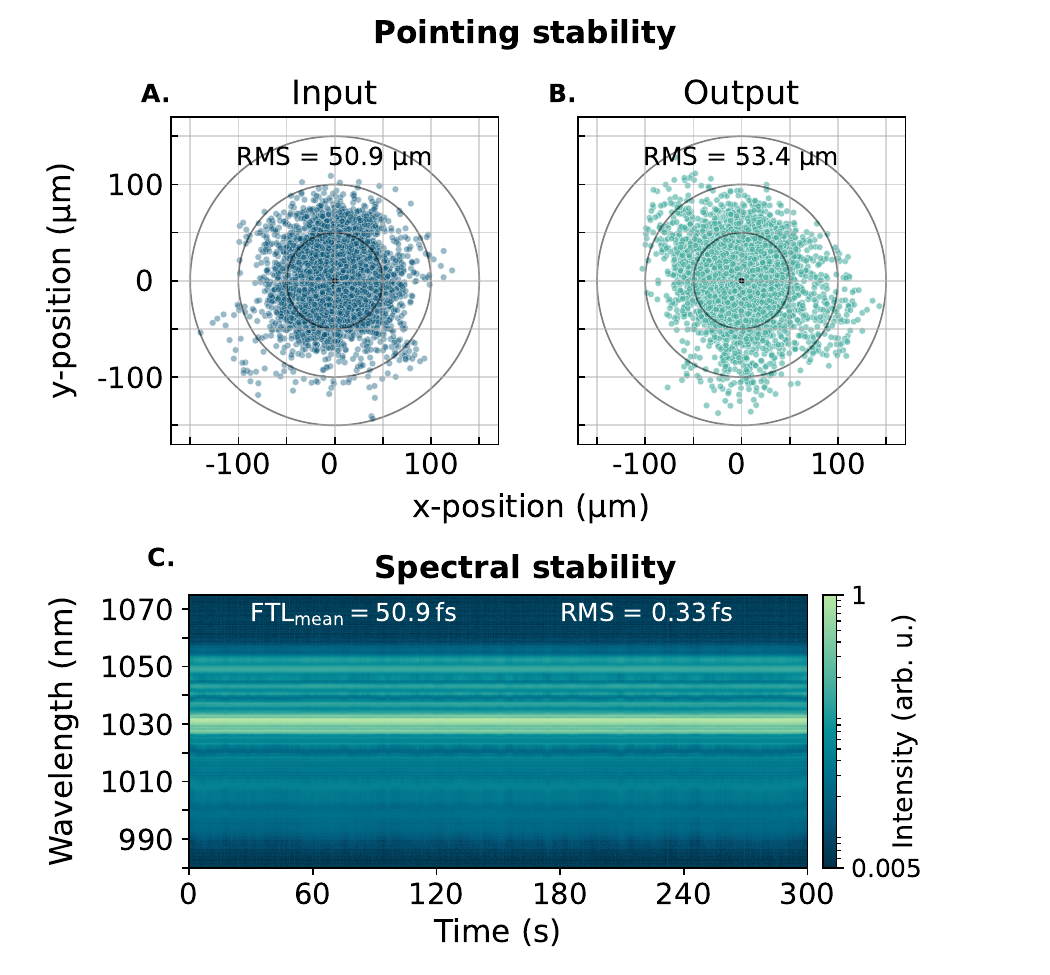}
    \caption{Beam pointing and spectral stability. (A) and (B) show scatter-plots of the near-field (NF) pointing of input (A) and output (B) beams. Measurements are taken separately, the data span for each measurement is 7.5 minutes and 4000 data points, with few minutes between measurements. In both measurements, the camera is placed at exactly \qty{42}{\cm} distance from the common reference point $z_0$ [Fig. \ref{fig:setup}]. Circles indicate deviations of 50, 100 and \qty{150}{\micro\m} from the mean value. (C) Measured broadened spectrum recorded over 5 minutes with 10 points per minute at full broadening (logarithmic scale). The RMS deviation of the FTL amounts to 0.65\%.
    }
    \label{fig:pointing}
\end{figure}

We further conduct spectral broadening experiments in 1 bar Krypton. 
Due to technical constraints originating from mechanical instabilities of the vacuum chamber, the chamber could only be operated at 1 bar pressure.
We measure the spectral broadening with \qty{6.5}{\milli\joule} pulse energy and the results are shown in Fig. \ref{fig:spectrum_results_Kr}, where we reach a FTL of approximately \qty{120}{\femto\s}.

The beam path inside the CMPC involves 550 reflections and a long propagation path of about \qty{190}{\m}. 
The beam is particularly sensitive to displacements of the folding mirrors PM1 and PM2 [Fig. \ref{fig:setup}], where most of the reflections take place. 
Thus, it is important to characterize the pointing stability of the beam. 
Figures \ref{fig:pointing}(A) and \ref{fig:pointing}(B)  show the measured near-field pointing stability for both input and output beams. 
We conduct the measurements separately within a few minutes. The measurements show only a very slight decrease of near-field pointing stability of approximately 5\%. 
This demonstrates the spatial stability owing to the input-to-output imaging property of the CMPC, also known from standard MPCs. 
In addition, we measure the spectral stability of the system over 5 minutes [Fig. \ref{fig:pointing}(C)]. We measure RMS fluctuations of the FTL of \qty{0.33}{\femto\s}, with a mean value of \qty{50.9}{\femto\s}, resulting in a relative RMS deviation of 0.65\%, indicating that the spectral broadening is stable.
\begin{figure}[tb]
    \centering
    \includegraphics[width=\textwidth]{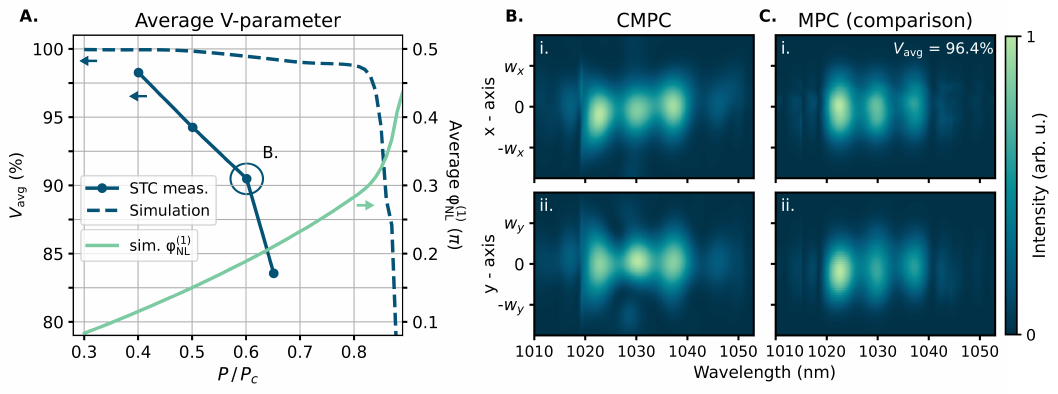}
    \caption{Measured spatio-spectral beam characteristics. (A) Measured and simulated average spectral overlap $V_{\mathrm{avg}}$ and simulated accumulated nonlinear phase $\phi_{\mathrm{NL}}^{(1)}$ as a function of peak power $P$ normalized to the critical power $P_{\mathrm{c}}$. (B) Example data set of a single measurement point, indicated by a circle in (A). (i) and (ii) show the spectral distribution of the pulse over the spatial $x$-direction, integrated over the $y$-direction and vice versa. (C) Corresponding STC traces for a standard MPC with similar broadening. $w_x$ and $w_y$ represent the $1/e^2$ beam radii.}
    \label{fig:STC}
\end{figure}

We further measure the beam quality parameter $\mathrm{M}^2$ of the input beam after the mode-matching telescope and the output beam. We observe a slight beam quality degradation from $\mathrm{M}^2=1.36$ at the input to $1.76$ at the output at full power.
As this observation may indicate spatio-temporal coupling (STC) effects, we decided to further investigate the spatio-spectral pulse characteristics using spatially resolved Fourier transform spectrometry \cite{Miranda2014}. 
The STC measurements provide information on the spatially-dependent spectral homogeneity of the pulse $V(x,y)$, which we calculate via the spectral overlap integral as defined in Eqs. (\ref{eq:Vparam})-(\ref{eq:Vparam2}) in the supplemental document.
An average spectral overlap $V_{\mathrm{avg}}$ is then computed by weighting the $V(x,y)$ with the fluence $F(x,y)$ obtained from the same dataset and averaging over an area defined by the measured $1/e^2$ diameter of the beam.
In air [Fig. \ref{fig:VCMPCair}], we measure $V_{\mathrm{avg}}=90.6\%$ at full power, confirming a slight  degradation of the spatio-spectral homogeneity compared to the input beam, which exhibits an almost perfect homogeneity of $V_{\mathrm{avg}}\approx 99\%$. 
We further investigate more generally how the spatio-spectral homogeneity behaves in atomic gases as a function of the input peak power $P$ of the pulse, compared to the critical power $P_c$ of the gas. 
The measurement is performed using Krypton as the nonlinear medium at 1 bar pressure and the same CMPC configuration is used as described in the beginning of this section.
We carry out STC measurements at four different pulse energies and thus four different values for $P/P_c$ and measure an STC trace for each point.
Figure \ref{fig:STC} summarizes the STC measurement results. 
From the first measurement point [Fig. \ref{fig:STC}(A)] at $P/P_c \approx 0.4$ to the third at $P/P_c \approx 0.6$, we observe a linear decline from 98.3\% to 90.5 \%. 
The fourth point at $P/P_c \approx 0.65$ exhibits a stronger deterioration. 
We compare our results with simulations considering again the measured input pulse as input for the simulation. 
Here we observe an onset of homogeneity reduction [Fig. \ref{fig:STC}(A)] at approximately $P/P_c = 0.55$ and an overall weaker deterioration compared to the measurements.
However, the simulations reproduce the measured behavior qualitatively including the onset of homogeneity degradation at around $P/P_c \approx 0.6$ well.
At $P/P_c \approx 0.8$, the simulations predict a beam collapse causing a sharp decline of $V_{\mathrm{avg}}$.
The faster degradation of the spatio-spectral homogeneity in the experimental data might be related to imperfect input beam characteristics in the experiment and/or possible spatial phase distortions arising due to many beam reflections on the CMPC mirrors. 

For an example point at $P/P_c \approx 0.6$, the spatio-spectral distribution obtained via an STC scan is shown in Fig. \ref{fig:STC}(B).
We compare the results with a spectrally broadened pulse from a standard gas-filled MPC with $N=17$, $k=16$ and $R=\qty{1}{\m}$ at similar spectral broadening characteristics.
The corresponding spatio-spectral distribution is shown in Fig. \ref{fig:STC}(C). Both measurements indicate quite homogeneous spatio-spectral characteristics and thus a similar spectral broadening performance for both MPC types, with $V_{\mathrm{avg}} = 96.4\%$ for the standard MPC and a slightly reduced $V_{\mathrm{avg}} = 90.5\%$ for the CMPC.

\section{Discussion}

The CMPC enables post-compression for high pulse energies at compact setup length, tunable via the folding ratio $\Gamma$ [Eq. (\ref{eq:EmaxCMPC})] while exhibiting excellent characteristics known from conventional MPCs. These include support for large compression factors, excellent beam quality and pointing properties as well as high transmission efficiency. 
In principle, $\Gamma$ can be increased to arbitrarily large values, enabling very compact systems.
There are, however, practical limitations.
The mirror size, in particular the length (corresponding to the x-dimension in Fig. \ref{fig:MPC_CMPC}) set a limit on $\Gamma$ (a mirror dimension estimate is provided in Eqs. (\ref{eq:ROC})-(\ref{eq:DCMPC}) in the supplemental document). 
Increasing $\Gamma$ also gives rise to a larger amount of total reflections.
This can lead to a decreased throughput and lower efficiency of the system.
In our experiment we measure a throughput of 89\% at 550 reflections, thus we can derive an average mirror reflectivity of $>$99.98\%.
Moreover, a large number of reflections can cause wavefront distortions, which we minimize by using mirrors with a high surface flatness of $\lambda/20$.
Another factor that can restrict energy scalability is the radius-of-curvature of the mirror substrate that can be manufactured. 

As discussed in section \ref{sec:concept} the CMPC scheme requires a regime where the beam inside the cell is loosely focused in order to enable beam folding on additional mirrors. 
This in turn fundamentally limits the amount of nonlinear phase $\phi_{\mathrm{NL}}^{(1)}$ which can be accumulated per pass to smaller values compared to a strongly focused geometry. 
In gas-filled MPCs, the accumulated Gouy phase per pass $\phi_{\mathrm{NL}}^{(1)}$ sets a theoretical limit on the nonlinear phase $\phi_{\mathrm{NL}}^{(1)}\leq\phi_{\mathrm{G}}^{(1)}$ \cite{Viotti2022}, which is directly related to the MPC geometry via $\phi_{\mathrm{G}}^{(1)} = \pi k/N$. 
In standard MPCs with $k/N \rightarrow 1$ this limit approaches $\phi_{\mathrm{NL}}^{(1)} = \pi$, whereas in the CMPC typically $k/N$ is lower.
In our experiment, the configuration is set to $N=11$ and $k=3$ and thus $\phi_{\mathrm{\mathrm{NL}}}^{(1)} \leq 0.27\pi$. 
Getting close to this limit, the peak power approaches the critical power of the medium $P_c$ and spatio-temporal couplings can become more pronounced leading to a sudden degradation of the spatio-spectral homogeneity.
However, this effect is not necessarily a limiting factor for spectral broadening in a CMPC.
The generally smaller $\phi_{\mathrm{NL}}^{(1)}$ in a CMPC can be compensated by simply employing more passes through the cell. For both MPC types, operation at $P < P_c$ ensures that excellent spatio-spectral and thus spatio-temporal pulse characteritics can be reached. 

\section{Outlook and conclusion}
\begin{figure}[b]
    \centering
    \includegraphics[width=0.55\textwidth]{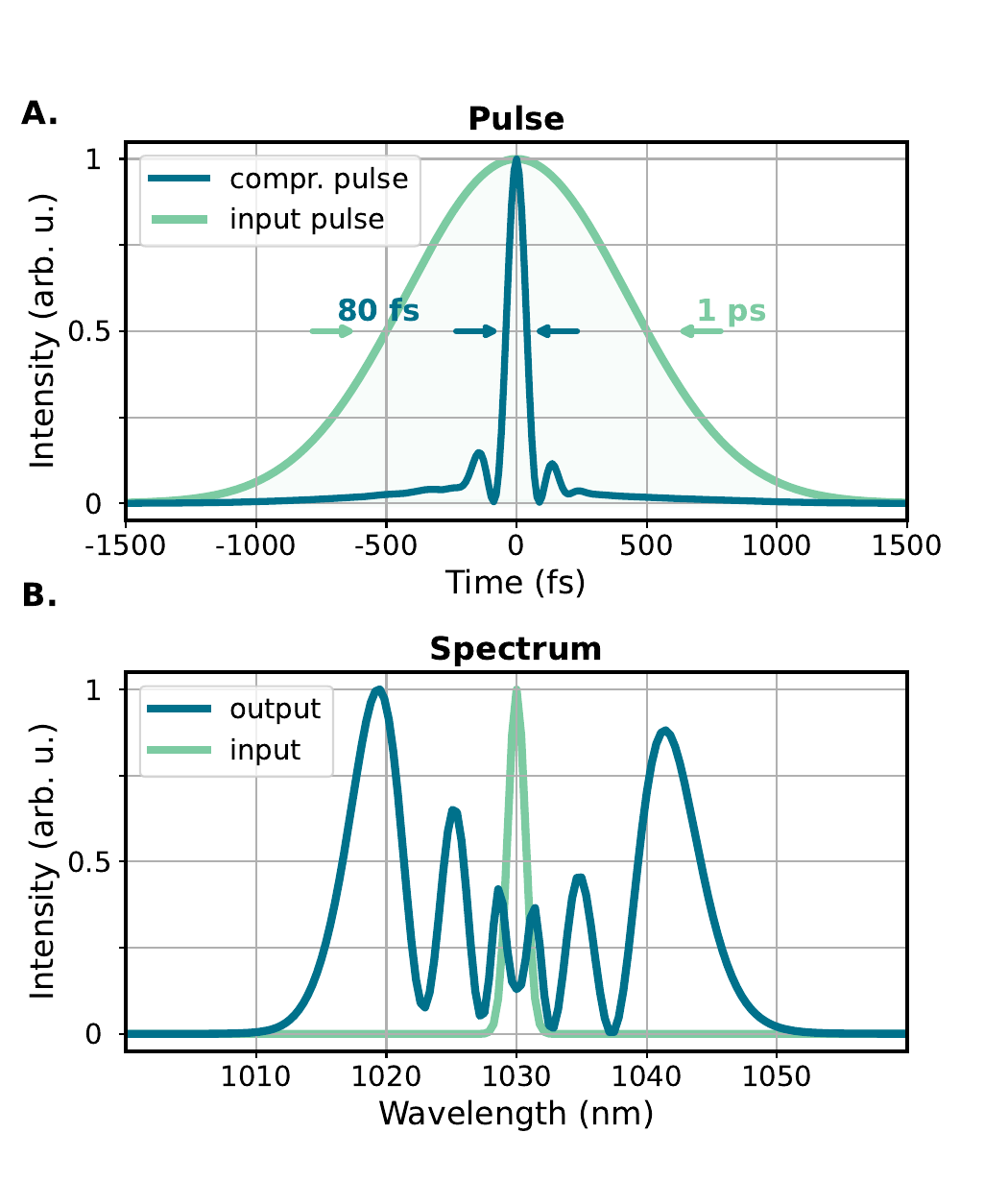}
    \caption{CMPC simulation for \qty{200}{mJ} pulses. (A) Input pulse and simulated compressed output pulse. (B) Input spectrum and simulated broadened spectrum.}
    \label{fig:sim_200mJ}
\end{figure}

In our proof-of-principle experiment we demonstrate post-compression of \qty{8}{\milli\joule} pulses in a compact setup.
However, we can further scale up the energy by increasing the mirror sizes as well as the radius-of-curvature $R$ of the mirrors.
As an outlook for high-energy CMPC operation, we simulate the post-compression of \qty{200}{\milli\joule}, \qty{1}{\pico\second} pulses at \qty{1030}{\nano\meter} central wavelength in a CMPC using \qty{50}{\milli\bar} of Argon as the nonlinear medium.
We show that post-compression down to \qty{80}{\femto\second} can be achieved, with a transmission of almost 80\%.
Figure \ref{fig:sim_200mJ} summarizes the simulation results for this scenario, indicating excellent post-compression performance. 
Here, we assume mirror reflectivity of 99.98\% and consider the influence of the mirror on the spectral phase of the pulse. 
We include the effects of all reflections on the mirrors in the simulations, which amount to $2N\Gamma = 1170$.
Using focusing mirrors with $R=\qty{300}{\m}$ and folding mirrors of $\qty{35}{\cm} \times \qty{35}{\cm}$, a folding ratio of $\Gamma = 39$ is achievable, assuming a clear aperture of $5$ times the beam radius $w_{\mathrm{m}}$ (see supplemental document section (S2)).
In this case an effective length of the setup of \qty {2.5}{\m} can be reached. 
The calculated fluence on the mirrors is kept at roughly of \qty{0.5}{\joule\per\cm^2}, including nonlinear self-focusing effects and nonlinear mode-matching taking into account Kerr lensing in the cell \cite{Hanna2021}.

In conclusion, we introduce a novel multi-pass cell scheme enabling post-compression of high-energy laser pulses in a compact setup. 
The CMPC enables tuning and down-scaling of the setup size via beam folding using additional planar mirrors, using weakly focused cell modes. 
Instead of increasing the length of the setup, the folding ratio $\Gamma$ acts as the energy scaling parameter. We demonstrate post-compression in air from \qty{1.1}{\pico\second} down to \qty{51}{\femto\second} in a CMPC with an effective length of \qty{45}{\cm} and a folding ratio $\Gamma = 25$, while keeping the fluence comparable to a standard MPC supporting the same pulse energy but requiring around \qty{2}{\m} cell length. 
Further up-scaling options promise post-compression of pulses with an energy of \qty{100}{\milli\joule} and beyond in a table-top setup.

\section*{Acknowledgement} 
We acknowledge Deutsches Elektronen-Synchrotron DESY (Hamburg, Germany), the Helmholtz Institute Jena (Jena, Germany), members of the Helmholtz Association HGF for support and/or the provision of experimental facilities. We further acknowledge the Helmholtz-Lund International Graduate School (project no. HIRS-0018) for funding and support as well as Vetenskapsrådet/Swedish Research Council (grant no. 2022-03519).

\section*{Disclosure} 
The authors declare no conflict of interest.

\section*{Data availability} 
Data underlying the results presented in this paper are not publicly available at this time but may be obtained from the authors upon reasonable request.

\newpage
\newpage
\title{Compact, folded multi-pass cells for energy scaling of post-compression: supplemental document}

\section*{S1. $\ \ $  Nonlinear Pulse Propagation Model}

The pulse propagation model used for simulations in this work is presented here. We use single-atomic gases as well as ambient air, which mainly consists of molecular gases (N$_2$ and O$_2$), as the nonlinear media. In order to conduct simulations, we thus need to take into account time-dependent nonlinear $3^{\mathrm{rd}}$-order effects, which stem from coupling of the electric field to the rotational states of the gas molecules \cite{Lin1976,Langevin2019}. The full equation for the propagation model used in this work can be written in the frequency and spatial frequency domain as:
\begin{align}
    \dv{E(k_x,k_y,\omega)}{z} = ik_z \, E(k_x,k_y,\omega) + P^{\mathrm{NL}}(k_x,k_y,\omega) \ ,
    \label{eq:nl_eq}
\end{align}
where $k_z = \sqrt{k^2(\omega) - k_x^2 - k_y^2} $, $k(\omega)=n(\omega)\omega/c_0$ with $\omega$ denoting the radial frequency, $k_x$ and $k_y$ the spatial wave-numbers, $c_0$ the speed of light in vacuum and $n(\omega)$ the refractive index. The second term $P^{\mathrm{NL}}$ in equation (\ref{eq:nl_eq}) contains all the nonlinear effects used in the model. Here we include the Kerr-effect via the nonlinear refractive index $n_2$ up to its first order derivative, as well as the single damped-oscillator model for the molecular response as described in references \cite{Langevin2019,Couairon2011}. The gas-specific single damped-oscillator model is described by the damping time $\Gamma$, the frequency $\Lambda$, the Raman-Kerr nonlinear refractive index $n_2^{\mathrm{R}}$ and the Raman-Kerr fraction $f_{\mathrm{R}}$.
In space and time domain, the nonlinear polarization $P_{\mathrm{NL}}$ can be written as:
\begin{align}
  P^{\mathrm{NL}}(x,y,t) = \  &\frac{\pi \varepsilon_0 c_0 n(\lambda)}{\lambda}\, \bigg[ \left(n_2 |E|^2 - i\left(\frac{\lambda}{2\pi} \frac{n_2}{c_0} + \dv{n_2}{\omega}\right)\left( 2 E^* \dv{E}{t} +  E \dv{E^*}{t} \right) \label{eq:Pnl1} \right) (1-f_{\mathrm{R}}) \\[6pt]
  &+\ f_{\mathrm{R}} \ n_2^{\mathrm{R}} \ \frac{\Gamma^2/4 + \Lambda^2}{\Lambda} \, \mathrm{Im}\Bigl\{\mathrm{e}^{-(\Gamma/2 - i\Lambda) t} \ \int_{-\infty}^{\infty}\mathrm{e}^{(\Gamma/2 - i\Lambda) t^\prime} |E(t^\prime)| \, \mathrm{d}t^\prime \Bigr\} \bigg] \ ,
  \label{eq:Pnl2}
\end{align}
where $E = E(x,y,t)$ is the electric field of the pulse, $E^*$ the complex conjugate of $E$, $\varepsilon_0$ the dielectric constant, $n(\lambda)$ the refractive index and $n_2$ the nonlinear refractive index. The first derivative $\mathrm{d}n_2/\mathrm{d}\omega$ is determined using the scaling formula described in reference \cite{Bree2010} (Eq. (12)). 
Equation (\ref{eq:Pnl2}) describes the delayed molecular response of the medium. For the case of 1 bar of air, we use $\Lambda = \qty{12}{\tera\hertz}$, $\Gamma = \qty{10}{\tera\hertz}$, $f_{\mathrm{R}} = 0.6$, which we extract from reference \cite{Langevin2019}, as well as the Raman-Kerr nonlinear refractive index $n_2^{\mathrm{R}} = \qty{42e-24}{\m^2\per\watt}$ \cite{Nibbering1997}. We further use $n_2 = \qty{8e-24}{\m^2\per\watt}$ \cite{Zahedpour2015} for air and $n_2 = \qty{24e-24}{\m^2\per\watt}$ for krypton \cite{Wahlstrand2012}.
We solve Eq. (\ref{eq:nl_eq}) using a radially symmetric (2+1)D split-step approach with $E=E(r,t)$ and the spatial coordinate $r = \sqrt(x^2+y^2)$, where in the spatial domain, the Fourier-transforms are replaced by Hankel-transforms\footnote[1]{In the simulations we use the "pyhank" package by Github user etfrogers for Hankel-transforms. }.

\section*{S2. $\ \ $  Calculation of CMPC mirror dimensions}
\begin{figure}[htb]
  \centering
  \includegraphics[width=0.6\textwidth]{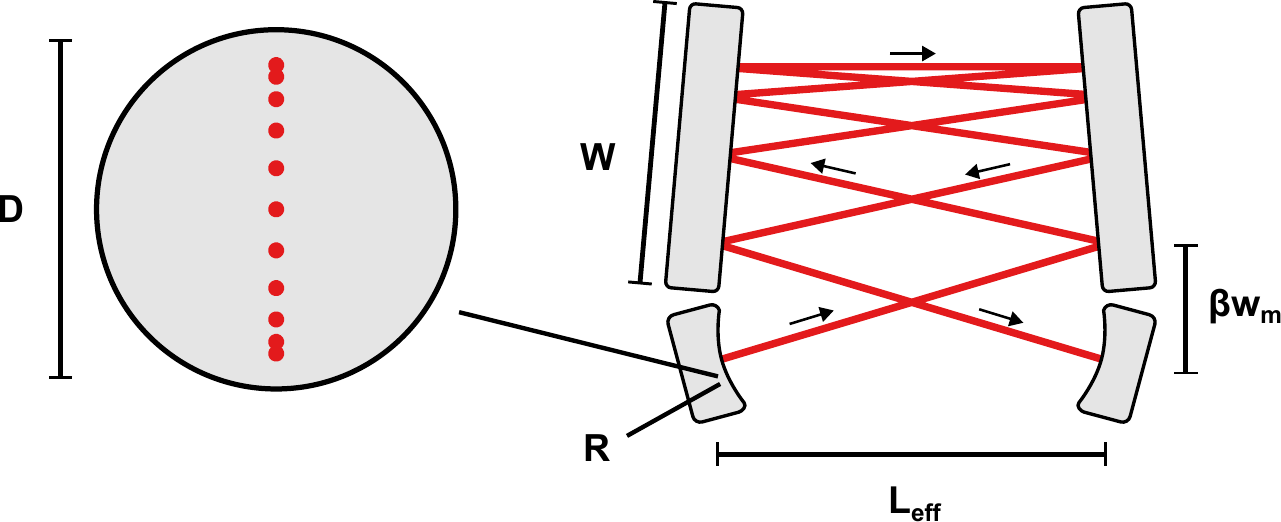}
  \caption{CMPC geometry.}
  \label{fig:CMPC_geometry}
\end{figure}

We here provide some useful equations describing the required minimum mirror dimensions for the CMPC. 
We consider a pulse energy $E$, a threshold fluence $F_{\mathrm{th}}$ as well as the CMPC configuration with $k$ and $N$.
We determine the focusing mirror radius-of-curvature $R$, the diameter of the focusing mirrors D (which is the same as the height of the planar mirrors) and the width of the planar folding mirrors W. The corresponding dimensions are shown in Fig. \ref{fig:CMPC_geometry}.
In order to calculate $R$, we use the equation for the fluence in the focus $F_0 = 4E/(\lambda R \sin{(\pi k/N)})$ \cite{Heyl2022}, set $F_0 \leq F_{\mathrm{th}}$ and re-arrange such that
\begin{align}
  R \ \geq \ \frac{4E}{\lambda F_{\mathrm{th}}} \, \frac{1}{\sin{(\pi k/N)}} \ .
  \label{eq:ROC}
\end{align}
To find out the width of the planar folding mirrors $W$, we need to ensure that the beam at any reflection has sufficient free aperture. 
For this, we define a factor $\beta$, where $\beta w_{\mathrm{m}}$ is the distance between the spot on the focusing mirror to the first reflection on the folding mirror [Fig. \ref{fig:CMPC_geometry}], and $w_{\mathrm{m}} = (R\lambda/\pi \cdot \tan{(\pi k/2N)})^{1/2}$ \cite{Heyl2022} is the $1/e^2$ beam radius on the focusing mirror.
With some basic geometric considerations, we arrive at:
\begin{align}
    W\, \geq \frac{\Gamma}{4}\,\beta\, \sqrt{\frac{4 E}{\pi F_{\text{th}}} \frac{\tan{(\pi k/2N)}}{\sin(\pi k/N)}},
    \label{eq:WCMPC}
\end{align}
defining the minimum width of the planar folding mirrors.
Here, $\Gamma$ is the folding ratio.
We typically choose a value $\beta = 5$ for our size estimations.
The height of the mirrors, or equivalently, the minimal diameter of the focusing mirrors $D$ can be calculated via:
\begin{align}
  D\, \geq \frac{(2N+1)\beta}{\pi^{3/2}} \sqrt{R\lambda\tan{(\pi k /2N)}}  \ .
  \label{eq:DCMPC}
\end{align}

\section*{S3. $\ \ $  Spectral homogeneity calculation and experimental data}

The spatio-spectral homogeneity, expressed as the $x$ and $y$-dependent spectral overlap $V(x,y)$ as it is used in reference \cite{Pfaff2023}, is calculated using the overlap integral
\begin{align}
    V(x,y) = \frac{\left[\int{I_0(\lambda)\,I(\lambda,x,y)\,\mathrm{d}\lambda}\right]^2}{\int{I_0^2(\lambda)\,\mathrm{d}\lambda} \, \int{\,I^2(\lambda,x,y)\,\mathrm{d}\lambda}} \times 100 \ ,
    \label{eq:Vparam}
\end{align}
where $\lambda$ is the wavelength, $I$ the spectral intensity and $I_0$ the spectral intensity at $(x,y) = (0,0)$. 
The averaged spectral homogeneity is then calculated using the average of $V(x,y)$ weighted with the wavelength-integrated intensity $F(x,y) = \int{I(\lambda,x,y)\,\mathrm{d}\lambda}$, yielding
\begin{align}
    V_{\mathrm{avg}} =  \frac{\int_{-w}^{w}{V(x,y) \, F(x,y) \mathrm{d}x \mathrm{d}y}}{\int_{-w}^{w}{F(x,y) \mathrm{d}x \mathrm{d}y}} \ ,
    \label{eq:Vparam2}
\end{align}
where $w$ is the $1/e^2$ beam radius in $x$- and $y$-direction respectively.

\begin{figure}[b]
  \centering
  \includegraphics[width=0.8\textwidth]{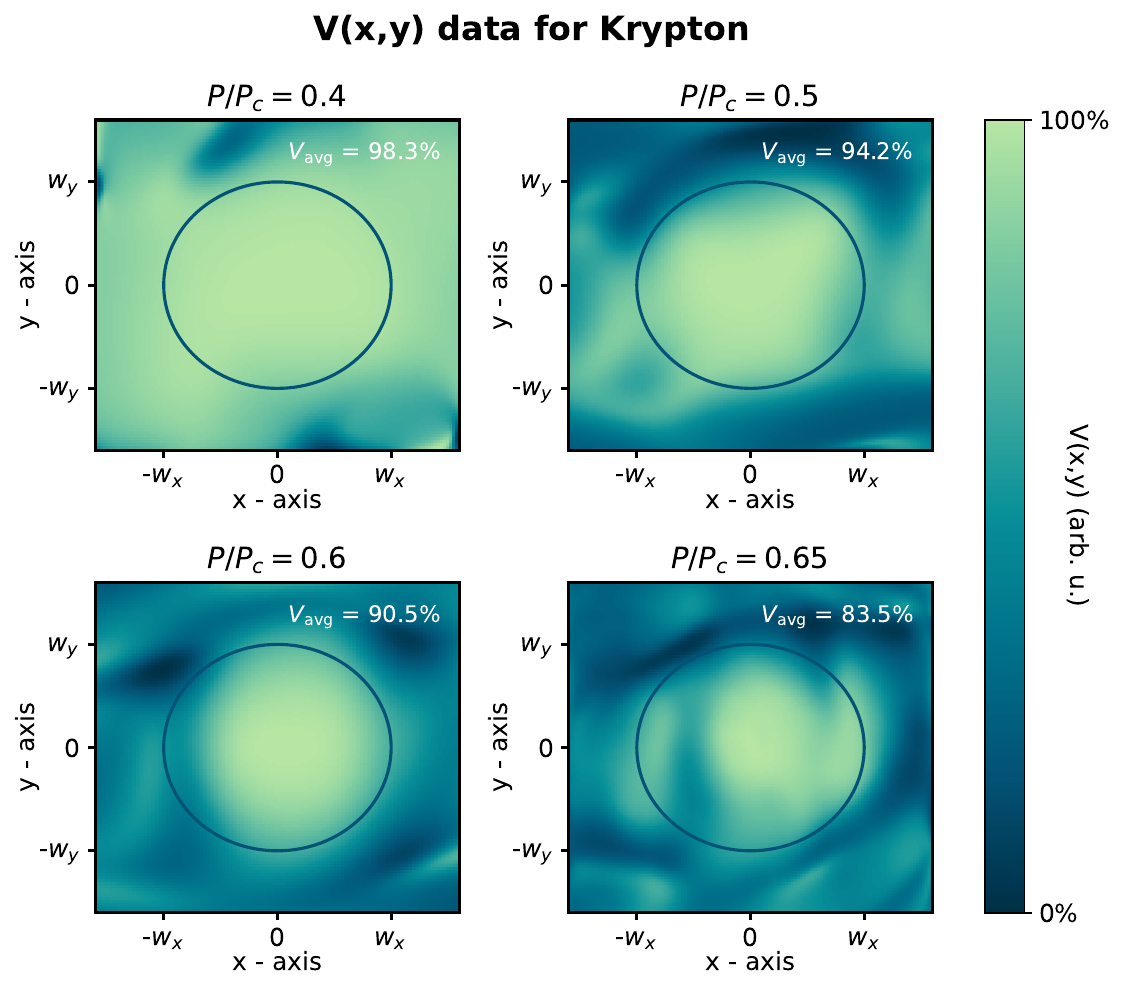}
  \caption{Spectral homogeneity $V(x,y)$ [Eq. (\ref{eq:Vparam})] for each $V_{\mathrm{avg}}$ data point shown in Fig. 8 in the main article. The circles indicate the area of integration for the calculation of $V_{\mathrm{avg}}$ [Eq. (\ref{eq:Vparam2})].}
  \label{fig:VCMPC}
\end{figure}

\begin{figure}[htb]
  \centering
  \includegraphics[width=0.45\textwidth]{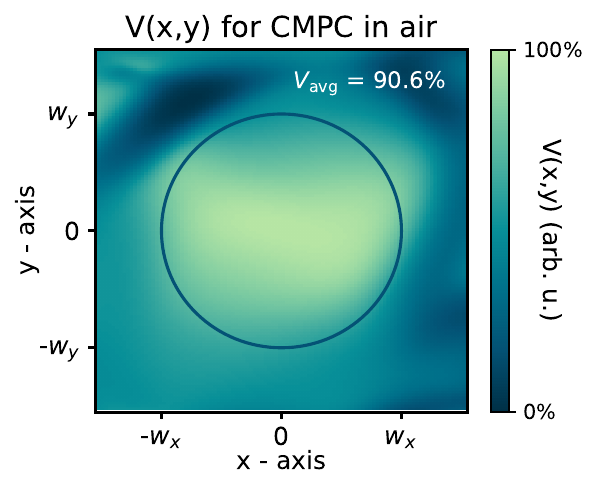}
  \caption{Spectral homogeneity $V(x,y)$ [Eq. (\ref{eq:Vparam})] for spectral broadening in air at \qty{8}{\milli\joule}. The circle indicates the area of integration for the calculation of $V_{\mathrm{avg}}$ [Eq. (\ref{eq:Vparam2})].}
  \label{fig:VCMPCair}
\end{figure}

\begin{figure}[htb]
  \centering
  \includegraphics[width=0.45\textwidth]{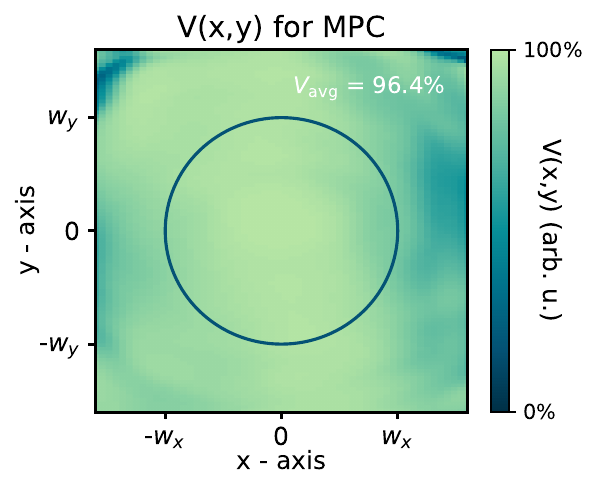}
  \caption{Spectral homogeneity $V(x,y)$ [Eq. (\ref{eq:Vparam})] for the MPC comparison measurement shown in Fig. 8(C) in the main article. The circle indicates the area of integration for the calculation of $V_{\mathrm{avg}}$ [Eq. (\ref{eq:Vparam2}]).}
  \label{fig:VMPC}
\end{figure}

In Figs. \ref{fig:VCMPC}-\ref{fig:VMPC} we display $V(x,y)$  for the measurements in krypton and air in the CMPC, as well as the comparison measurement conducted in a standard MPC.
Figure \ref{fig:VCMPC} shows $V(x,y)$ for each measurement point which is displayed in Figure (8)(A) in the main article.
In Fig. \ref{fig:VCMPCair}, the spectral homogeneity is shown for the air measurements, carried out at the same parameters as the main spectral broadening and post-compression measurements shown in Fig. 5 in the main article with 1 bar air and \qty{8}{\milli\joule} pulse energy.
Finally, Fig. \ref{fig:VMPC} shows $V(x,y)$ for the comparison measurement in a standard MPC.

\clearpage

\bibliography{literature/CMPC_library}

\end{document}